\documentclass[12pt]{article}
\usepackage{amssymb}

\newcommand{\be}{\begin{equation}}
\newcommand{\ee}{\end{equation}}
\newcommand{\bea}{\begin{eqnarray}}
\newcommand{\eea}{\end{eqnarray}}
\newcommand{\ba}{\begin{array}}
\newcommand{\ea}{\end{array}}
\newcommand{\bi}{\begin{itemize}}
\newcommand{\ei}{\end{itemize}}
\newcommand{\bn}{\begin{enumerate}}
\newcommand{\en}{\end{enumerate}}
\newcommand{\bc}{\begin{center}}
\newcommand{\ec}{\end{center}}
\renewcommand{\l}{\left}
\renewcommand{\r}{\right}

\newcommand{\ol}{\overline}

\newcommand{\nl}{\nonumber\\}

\begin{document}
\tolerance=100000
%\thispagestyle{empty}
%\setcounter{page}{0}

%\begin{flushright}
%{\tt hep-ph/0605094\\
%KIAS-TH-04XXX}
%\end{flushright}

\vspace*{\fill}

\begin{center}
{\Large \bf
New physics in $B \to \pi \pi$ and $B \to \pi K$ decays
}\\[3.cm]

{\large\bf Seungwon Baek\footnote{sbaek@cskim.yonsei.ac.kr}}
\\[7mm]

{\it
Department of Physics, Yonsei University, Seoul 120-749, Korea
}\\[10mm]
\end{center}

\vspace*{\fill}

\begin{abstract}
{\small\noindent
We perform a combined analysis of $B \to \pi \pi$ and $B \to \pi K$ decays
with the current experimental data. Assuming  $SU(3)$ flavor symmetry
and no new physics contributions to the topological amplitudes,
we demonstrate that the conventional parametrization in the
Standard Model (SM) does not describe
the data very well, in contrast with a similar analysis based on the
earlier data.
It is also shown that the introduction of smaller amplitudes and
reasonable $SU(3)$ breaking parameters does not improve the fits much.
Interpreting these puzzling behaviors in the SM
as a new physics (NP) signal, we study various
NP scenarios. We find that when a single NP amplitude dominates, the NP
in the electroweak penguin sector is
the most favorable. However, other NP solutions, such as
NP residing in the QCD-penguin sector and color-suppressed
electroweak penguin sector simultaneously, can also solve the puzzle.
}
\end{abstract}

\vspace*{\fill}

\begin{flushleft}
{\rm  April 2006} \\
\end{flushleft}

\newpage

\section{Introduction}
%B-factories have produced a lot of $B$-mesons, and the study of
%$B$-meson decay processes is entering the precision era.
In the Standard Model (SM), rare non-leptonic decays $B \to \pi\pi$
and $B \to \pi K$ provide valuable information on the inner angles of
the unitarity triangle of Cabbibo-Kobayashi-Maskawa (CKM) matrix, and have
been widely studied.
For this purpose the measurement of time-dependent CP-asymmetry
given by
\bea
\frac{\Gamma(\ol{B}(t)\to f_{\rm CP})-\Gamma(B(t)\to f_{\rm CP})}
     {\Gamma(\ol{B}(t)\to f_{\rm CP})+\Gamma(B(t)\to f_{\rm CP})}
  = A_{\rm CP} \cos(\Delta m t)
  + S_{\rm CP} \sin(\Delta m t)
\eea
is essential.
Here $A_{\rm CP}$ and $S_{\rm CP}$ represent direct and indirect
CP asymmetries, respectively.

Specifically, $B \to \pi\pi$ decays measure the angle $\alpha$ through
the isospin analysis~\cite{London-Gronau-1990}. The information on $\gamma$
can be obtained from $B \to \pi K$
data~\cite{Gronau:1994,Buras:2000,Imbeault:2003}.
In addition, if a new physics (NP) beyond the SM exists,
it can significantly affect these
processes by contributing to penguin amplitudes.
Therefore these decay processes are also a sensitive probe of
NP~\cite{Buras:EW,Baek:piK,Baek:pipi}.

Assuming i) $SU(3)$ flavor symmetry of strong interactions and
ii) smallness
of annihilation and exchange topologies, Buras, Fleischer, Recksiegel
and Schwab~\cite{Buras:EW}
concluded the $B \to \pi \pi$ and $B \to \pi K$ data strongly suggest
a NP in the electroweak penguin sector of
$B \to \pi K$ decay amplitudes. On the other hand,
Chiang, Gronau, Rosner and Suprun~\cite{Rosner:2004} demonstrated that
the $\chi^2$-fit to the same data
does not show any significant deviation from the SM.
It should be noted that Buras, {\it et.al.}~\cite{Buras:EW}
assumed that there is no significant NP contribution to
$B \to \pi\pi$ decays, and used some of $B \to \pi\pi$ data with small
experimental errors to predict the hadronic parameters of
$B\to \pi K$ amplitudes.
Chiang, {\it et.al}~\cite{Rosner:2004} included all the
available $B \to \pi \pi$ and $B \to \pi K$ data into their fit.

We considered $B \to \pi K$ data only~\cite{Baek:piK},
and showed that the SM fit faces some difficulties and
NP in the electroweak penguin sector is strongly favored
in accord with~\cite{Buras:EW}.
We concluded that the discrepancy between \cite{Buras:EW,Baek:piK}
and \cite{Rosner:2004} is due to the dilution of NP effects by including
all the data in \cite{Rosner:2004}.

In this paper, we perform $\chi^2$ fitting to the current data in the
SM and also in the presence of NP.
We show that even if following the approach of
Chiang, {\it et.al}~\cite{Rosner:2004},
{\it i.e.} the $\chi^2$ fitting with all the available data,
we get much worse $\chi^2$ fit
than in~\cite{Rosner:2004}.
We calculated $\Delta \chi^2$ --the contribution of
each data point to $\chi^2$ value--
to trace the source of this puzzling behavior.
%to find that the data, $B(B^0 \to \pi^0 K^0)$,
%$A_{CP}(B^+ \to \pi^+ K^0)$ and $S_{CP}(B^0 \to \pi^0 K^0)$,
%give the largest contributions.
To improve the fit in the SM, we introduce i) smaller amplitudes,
and/or ii) reasonable $SU(3)$-breaking effects to the fits.
It turns out that these corrections do not solve the
puzzle satisfactorily.

Interpreting these difficulties in the SM fits as a NP signal,
we introduce NP parameters, such as, a new weak phase in the
amplitudes.
We consider various NP scenarios.
We introduce three types of NP, NP in the
electroweak penguin, color-suppressed electroweak penguin and
QCD penguin.
When a single NP amplitude dominates, NP in the electroweak penguin
is the most favorable solution, supporting the findings in~\cite{Buras:EW}.
A given specific NP model, however, contributes to all the NP amplitudes
in general.
In light of this we also considered the possibility two or more
NP amplitudes are enhanced simultaneously.

The paper is organized as follows.
In Section~\ref{sec:SM} the SM fittings are considered.
In Section~\ref{sec:NP} we perform various NP fittings.
The conclusions and discussions are given in Section~\ref{sec:con}.

\section{SM fitting}
\label{sec:SM}
The topological amplitudes provide a
parametrization for non-leptonic $B$-meson decay processes
which is independent of theoretical models for the calculation
of hadronic matrix elements~\cite{GHLR}.
The decay amplitudes of $B\to \pi\pi$'s which are $\ol{b}\to \ol{d}q\ol{q}$
$(q=u,d)$ transitions at quark-level can be written as
\bea
\sqrt{2} A(B^+ \to \pi^+ \pi^0) &=& -\l(T + C + P_{\rm EW} + P^C_{\rm EW} \r),
\nl
A(B^0 \to \pi^+ \pi^-) &=& -\l(T + P + {2 \over 3} P^C_{\rm EW} + E + PA \r), \nl
\sqrt{2} A(B^0 \to \pi^0 \pi^0) &=& -\l(C - P +P_{\rm EW}
+ {1 \over 3}
P^C_{\rm EW} - E -PA\r).
\label{eq:pipi_full}
\eea
Here $T$, $C$, $P$, $P_{\rm EW}^{(C)}$, $E$ and $PA$ represent
tree, color-suppressed tree, QCD-penguin, (color-suppressed)
electroweak-penguin,
exchange and penguin annihilation diagrams, respectively.
Similarly, $B \to \pi K$ decays which are $\ol{b}\to \ol{s}q\ol{q}$
$(q=u,d)$ transitions at quark-level are described by
\bea
A(B^+ \to \pi^+ K^0) &=& P' -{1 \over 3} P^{'C}_{\rm EW} + A', \nl
\sqrt{2} A(B^+ \to \pi^0 K^+) &=& -\l(P' + T' + C' + P'_{\rm EW} +
{2 \over 3} P^{'C}_{\rm EW} +A' \r), \nl
A(B^0 \to \pi^- K^+) &=& - \l( P' + T' + {2 \over 3} P^{'C}_{\rm EW} \r), \nl
\sqrt{2} A(B^0 \to \pi^0 K^0) &=&
P' - C'- P'_{\rm EW}  -{1 \over 3} P^{'C}_{\rm EW},
\label{eq:piK_full}
\eea
where primes indicate $b \to s$ transition. The corresponding
decay amplitudes for the CP-conjugate modes can be obtained by
changing the sign of weak phases while keeping CP-conserving strong
phases unchanged.

We can further decompose the QCD penguin diagrams, $P$ and $P'$, depending on the
quarks running inside the loop,
\bea
P &=& V_{ud} V^*_{ub} P_u + V_{cd} V^*_{cb} P_c + V_{td} V^*_{tb} P_t \nl
  &=& V_{ud} V^*_{ub} (P_u-P_c) + V_{td} V^*_{tb} (P_t-P_c) \nl
  &\equiv& P_{uc} e^{i\gamma} + P_{tc} e^{-i\beta}, \nl
P' &=& V_{us} V^*_{ub} P'_u + V_{cs} V^*_{cb} P'_c + V_{ts} V^*_{tb} P'_t \nl
  &=& V_{us} V^*_{ub} (P'_u-P'_c) + V_{ts} V^*_{tb} (P'_t-P'_c) \nl
  &\equiv& P'_{uc} e^{i\gamma} - P'_{tc},
\eea
where we have used the unitarity relation for CKM matrix elements
and explicitly written the weak phase dependence for the amplitudes.
These notations and conventions will be used throughout the paper.
We can estimate the relative sizes of the amplitudes based on
the color-, CKM-, and loop-factors,
\bea
\begin{array}{ccc}
   & B \to \pi\pi & B \to \pi K \\
\hline
O(1)  & |T|  & |P'_{tc}| \\
O(\bar{\lambda}) &  |C|, |P| & |T'|, |P'_{\rm EW}| \\
O(\bar{\lambda}^2) &  |P_{\rm EW}|
 & |C'|, |P'_{uc}|,|P^{'C}_{\rm EW}| \\
O(\bar{\lambda}^3) &  |P^C_{\rm EW}| & |A'| \\
O(\bar{\lambda}^4) &  |E|,|PA| & \\
\label{eq:hierarchy}
\end{array}
\eea
where $\bar{\lambda}$ is expected to be order of $0.2\sim 0.3$.
We will call the decay amplitudes parameterized as in (\ref{eq:pipi_full})
and (\ref{eq:piK_full}) and the hierarchy in (\ref{eq:hierarchy})
the conventional parametrization in the SM.

The decay amplitudes containing only dominant terms, $T^{(')}$,
$C^{(')}$~\footnote{We include $C'$ to the amplitudes,
although  according to~(\ref{eq:hierarchy}) it is subdominant.
Otherwise, we get extremely poor fit~\cite{Baek:piK}. We hope
that this problem will be solved within the SM framework.}
,$P_{tc}^{(')}$ and $P'_{\rm EW}$ are given by~\footnote{
We may also think of $T^{(')}$ and $C^{(')}$ as
$T^{(')}+P^{(')}_{uc}$ and $C^{(')}-P^{(')}_{uc}$, respectively.
See \cite{Rosner:2004,Baek:piK}, for details.
}
\bea
\sqrt{2} A(B^+ \to \pi^+ \pi^0) &=& -\l(T + C \r)  e^{i\gamma},
\nl
A(B^0 \to \pi^+ \pi^-) &=& -\l(T e^{i\gamma} + P e^{-i\beta} \r), \nl
\sqrt{2} A(B^0 \to \pi^0 \pi^0) &=& -C^{i\gamma} + P e^{-i\beta}, \nl
A(B^+ \to \pi^+ K^0) &=& -P', \nl
\sqrt{2} A(B^+ \to \pi^0 K^+) &=& P' - T'e^{i\gamma}  - C'e^{i\gamma}
 - P'_{\rm EW} , \nl
A(B^0 \to \pi^- K^+) &=& P' - T'e^{i\gamma} , \nl
\sqrt{2} A(B^0 \to \pi^0 K^0) &=&
-P' - C' e^{i\gamma} -P'_{\rm EW}.
\label{eq:all}
\eea
Here, we have written $P^{(')}_{tc}$ as $P^{(')}$ for the simplicity
of notations.

\begin{table}[tbh]
\center
\begin{tabular}{cccc}
\hline
\hline
Mode & $BR[10^{-6}]$ & $A_{\rm CP}$ & $S_{\rm CP}$ \\ \hline
$B^+ \to \pi^+ \pi^0$ & $5.5 \pm 0.6$ & $0.01 \pm 0.06$ & \\
$B^0 \to \pi^+ \pi^-$ & $5.0 \pm 0.4$ & $0.37 \pm 0.10$ & $-0.50 \pm 0.12$ \\
$B^0 \to \pi^0 \pi^0$ & $1.45 \pm 0.29$ & $0.28 \pm 0.40$ & \\
\hline
$B^+ \to \pi^+ K^0$ & $24.1 \pm 1.3$ & $-0.02 \pm 0.04$ & \\
$B^+ \to \pi^0 K^+$ & $12.1 \pm 0.8$ & ~~$0.04 \pm 0.04$ & \\
$B^0 \to \pi^- K^+$ & $18.9 \pm 0.7$ & $-0.115 \pm 0.018$ & \\
$B^0 \to \pi^0 K^0$ & $11.5 \pm 1.0$ & ~~$0.02 \pm 0.13$ &
~~$0.31 \pm 0.26$ \\
\hline
\hline
\end{tabular}
\caption{The current experimental data for CP averaged branching ratios ($BR$),
direct CP asymmetries ($A_{\rm CP}$)
and indirect CP asymmetries ($S_{\rm CP}$)
for $B \to \pi \pi$ and $B \to \pi K$ decays filed by
HFAG~\cite{HFAG}.}
\label{tab:data}
\end{table}

The current experimental data for the CP-averaged branching ratio ($BR$),
the direct CP-asymmetry ($A_{\rm CP}$) and the indirect CP-asymmetry
($S_{\rm CP}$) are shown in Table~\ref{tab:data}.
It immediately shows some puzzling behaviors which are difficult
to understand if we believe (\ref{eq:hierarchy}) and (\ref{eq:all}).
Firstly, (\ref{eq:hierarchy}) suggests that $BR(B^0 \to \pi^0\pi^0)$
should be about 3 times lower than the data
($B \to \pi\pi$ puzzle)~\cite{Baek:pipi,B2pipi_puzzle}.
Secondly, the ratios
\bea
  R_c &\equiv& \frac{2 BR(B^+ \to \pi^0 K^+)}{BR(B^+ \to \pi^+ K^0)}, \nl
  R_n &\equiv& \frac{BR(B^0 \to \pi^- K^+)}{2 BR(B^0 \to \pi^0 K^0)}
\label{eq:Rc_Rn}
\eea
should equal to a good approximation.
However, the data shows
%$(1.00 \pm 0.09)-(0.82 \pm 0.08) = 0.18 \pm 0.12$ ($1.5 \sigma$)
about $1.5 \sigma$ difference. We should say that
this so-called $R_c/R_n$ problem is not so statistically significant
now.
Thirdly, we expect from (\ref{eq:all}) that
\bea
A_{\rm CP}(B^+ \to \pi^0 K^+)
\approx A_{\rm CP}(B^0 \to \pi^- K^+).
\label{eq:direct_CP}
\eea
The data deviate from this relation by about $2.7 \sigma$ level.
Finally, the dominant terms in $S_{\rm CP}(B^0 \to \pi^0 K^0)$ gives
$\sin 2\beta$ which is quite precisely measured from
$b \to s c\ol{c}$ modes to be $\sin 2\beta=0.685 \pm 0.032$~\cite{HFAG}.
The current data shows about $1.43 \sigma$ difference.
These last three are usually called
``$B \to \pi K$ puzzle''~\cite{B2piK_puzzle}.

Assuming the exact  $SU(3)$ flavor symmetry, we can relate
the topological amplitudes of $B \to \pi \pi$ decays
to the corresponding amplitudes of $B \to \pi K$ decays as follows:
\bea
 {T \over T'} &=& {C \over C'} = {V_{ud} \over V_{us}}, \nl
 {P \over P'} &=& \l|V_{td} \over V_{ts} \r|.
\label{eq:su3}
\eea
In addition, it is known that the Wilson coefficients for the
electroweak penguins
$c_7$ and $c_8$ are much smaller than $c_9$ and $c_{10}$~\cite{Buchalla:1995}
in the SM, which
leads to a relation between the electroweak penguin diagrams
and trees in the $SU(3)$-limit~\cite{GPY},
\bea
P'_{\rm EW} &=& {3 \over 4} {c_9 + c_{10} \over c_1 + c_2} R(T' + C')
+ {3 \over 4} {c_9 - c_{10} \over c_1 - c_2} R(T' - C'), \nl
P^{'C}_{\rm EW} &=& {3 \over 4} {c_9 + c_{10} \over c_1 + c_2} R(T' + C')
- {3 \over 4} {c_9 - c_{10} \over c_1 - c_2} R(T' - C').
\label{eq:GPY}
\eea
Here, $R$ is given by a combination of CKM matrix elements,
\bea
R = \l|V_{ts} V^*_{tb} \over V_{us} V^*_{ub}\r| =
{1 \over \lambda^2} {\sin(\beta+\gamma) \over \sin\beta}.
\eea

\begin{table}[t]
\center
\begin{tabular}{cccc}
\hline
\hline
 & SM fit I   & SM fit II & SM fit III\\
\hline
$\chi_{\rm min}^2/dof~(\mbox{quality of fit})$ & $18.8/10~(4.3\%)$
& $0.62/5~(99\%)$ & $16.4/8~(3.7\%)$\\
\hline
$\gamma$ & $69.4^\circ \pm 5.8^\circ$ & $73.2^\circ \pm 5.2^\circ$
         & $70.6^\circ \pm 5.2^\circ$ \\
$|T'|$~(eV) & $5.22 \pm 0.26$ & $5.26 \pm 0.27$
         & $6.59 \pm 0.29$ \\
$\delta_{T'}$ & $28.3^\circ \pm 4.8^\circ$ & $29.9^\circ \pm 5.3^\circ$
         & $25.0^\circ \pm 6.8^\circ$ \\
$|C'|$~(eV) & $3.82 \pm 0.48$ & $3.20 \pm 0.52$
         & $4.02 \pm 0.44$ \\
$\delta_{C'}$ & $-40.2^\circ \pm 9.7^\circ$ & $-14.3^\circ \pm 18.0^\circ$
              & $-47.0^\circ \pm 10.5^\circ$ \\
$|P'|$~(eV) & $48.9 \pm 0.7$ & $47.2 \pm 1.7$
            & $36.9 \pm 4.8$\\
%$\delta_{P'}$ & 0 (fixed) & 0 (fixed)\\
$|P'_{uc}|$~(eV) & - & - & $24.1 \pm 6.4$ \\
$\delta_{P'_{uc}}$~(eV) & - & - & $178^\circ \pm 2^\circ$ \\
\hline
\hline
\end{tabular}
\caption{Results for ``SM fit I'', ``SM fit II'' and ``SM fit III''.
See the text for details.}
\label{tab:SM-fit}
\end{table}

Using these $SU(3)$ relations, we have 6 parameters to fit in (\ref{eq:all}):
$|T'|$, $|C'|$, $|P'|$, two relative strong phases and $\gamma$ ({\bf SM fit I}).
Now we can perform the fit to the current experimental data which are given
in Table~\ref{tab:data}. Since in $P^{'C}_{\rm EW}$ is neglected
in $(\ref{eq:all})$, we use for this fit
\bea
  P'_{\rm EW} ={3 \over 2} {c_9 + c_{10} \over c_1 + c_2} R(T' + C'), \quad P^{'C}_{\rm EW} =0.
\eea
The inner angle $\beta$ of the unitarity triangle
is strongly constrained and is given by $\sin 2\beta=0.725 \pm 0.018$, so
we fixed $\beta = 23.22^\circ$.
Other parameters used as inputs for the fit are as follows:
$\lambda=0.226$, $c_1 = 1.081$, $c_2=-0.190$, $c_9 = -1.276 \alpha_{\rm em}$,
$c_{10} = 0.288 \alpha_{\rm em}$.
The result for this fit  is shown in
Table~\ref{tab:SM-fit}.~\footnote{We have also checked the case where
$P'_{EW} \approx -{3 \over 2} {(c_9+c_{10}) \over (c_1+c_2)} \; q_{\rm EW} \; R \;T'$
and $q_{\rm EW}$
is fitted as in~\cite{Rosner:2004}.
In this case we obtained $q_{\rm EW}=0.36 \pm 0.33$ which
is far away from the SM expectation $\delta_{\rm EW}=1$.
Therefore we used the exact $SU(3)$ relation (\ref{eq:GPY}) for this fit.}
It can be seen that the SM with exact $SU(3)$ symmetry gives
$\chi^2_{\rm min}/dof=18.8/10 (4.3\%)$, which is quite a poor fit.
To trace the observables which make the fit poor, we list
$\Delta \chi_{\rm min}^2$--the contribution of each
data point to the $\chi_{\rm min}^2$ in Table~\ref{tab:dchi2}.
From Table~\ref{tab:dchi2} we can see that the observables,
$BR(B^0 \to \pi^0 K^0)$, $A_{\rm CP}(B^+ \to \pi^0 K^+)$ and
$S_{\rm CP}(B^0 \to \pi^0 K^0)$ which caused the $B \to \pi K$
puzzles are exactly those with large $\Delta \chi^2_{\rm min}$.
They are about $1.8\sigma \sim 2.2 \sigma$ away from the best fit values.

An alternative way to see the discrepancy between the SM and the experiments
is to remove the observables which give large
$\Delta \chi^2_{\rm min}$ from the fit and predict them from the fitted
parameters of the remaining observables. For example, we dropped
the data for $BR(B^+ \to \pi^+ K^0)$, $BR(B^0 \to \pi^- K^+)$, $BR(B^0 \to \pi^0 K^0)$,
$A_{\rm CP}(B^+ \to \pi^0 K^+)$ and $S_{\rm CP}(B^0 \to \pi^0 K^0)$
whose $\Delta \chi_{\rm min}^2$'s from ``SM fit I'' are greater than 1 from
the $\chi^2$ fitting.
The results for this
approach ({\bf SM fit II}) are shown in the 2nd
columns of Tables~\ref{tab:SM-fit} and~\ref{tab:dchi2}.
From Table~\ref{tab:SM-fit}, we can see the quality of fitting has improved
dramatically while the values of parameters are consistent with
those of ``SM fit I''. Also Table~\ref{tab:dchi2} shows that all the
observables considered are excellently described by the SM parametrization.
Now we can predict the omitted observables from the fitted values
in Table~\ref{tab:SM-fit}.
The predictions (deviation from the best fit values) for
$BR(B^+ \to \pi^+ K^0)$, $BR(B^0 \to \pi^- K^+)$, $BR(B^0 \to \pi^0 K^0)$,
$A_{\rm CP}(B^+ \to \pi^0 K^+)$ and $S_{\rm CP}(B^0 \to \pi^0 K^0)$
are
$21.0 \pm 0.6~(2.1 \sigma)$, $18.5 \pm 0.4~(0.46 \sigma)$,
$8.17 \pm 0.16~(3.3 \sigma)$, $-0.065 \pm 0.002~(2.6 \sigma)$ and
$0.81 \pm 0.0001~(1.9 \sigma)$, respectively.
These deviations imply that the $B \to \pi K$ puzzles are more
serious than the estimations given below (\ref{eq:Rc_Rn})
and below (\ref{eq:direct_CP}).

\begin{table}[t]
\center
\begin{tabular}{cccc}
\hline
\hline
Observable & SM fit I  & SM fit II & SM fit III\\
\hline
$BR(B^+ \to \pi^+ \pi^0)$ & 0.67 & 0.00074 & 0.24\\
$BR(B^0 \to \pi^+ \pi^-)$ & 0.31 & 0.0086 & 0.068\\
$BR(B^0 \to \pi^0 \pi^0)$ & 0.83 & 0.0013 & 0.46\\
$BR(B^+ \to \pi^+ K^0)$ & 1.2 & - & 0.17\\
$BR(B^+ \to \pi^0 K^+)$ & 0.025 & 0.00679 & 0.59\\
$BR(B^0 \to \pi^- K^+)$ & 1.3 &  - & 0.98\\
$BR(B^0 \to \pi^0 K^0)$ & 4.8 & -& 1.4\\
\hline
$A_{\rm CP}(B^+ \to \pi^+ \pi^0)$ & 0.028 & 0.028 &0.028\\
$A_{\rm CP}(B^0 \to \pi^+ \pi^-)$ & $9.9 \times 10^{-5}$ & 0.10 & 1.1\\
$A_{\rm CP}(B^0 \to \pi^0 \pi^0)$ & 0.50 & 0.029 & 0.40\\
$A_{\rm CP}(B^+ \to \pi^+ K^0)$ & 0.25 & 0.25 & 0.18\\
$A_{\rm CP}(B^+ \to \pi^0 K^+)$ & 3.1 & - & 3.1\\
$A_{\rm CP}(B^0 \to \pi^- K^+)$ & 0.68 & 0.044 & 1.2\\
$A_{\rm CP}(B^0 \to \pi^0 K^0)$ & 0.33 & 0.15 & 0.49\\
\hline
$S_{\rm CP}(B^0 \to \pi^+ \pi^-)$ & 0.85 & 0.00013 & 0.19\\
$S_{\rm CP}(B^0 \to \pi^0 K^0)$ & 3.9 & -& 5.9\\
\hline
\hline
\end{tabular}
\caption{$\Delta \chi_{\rm min}^2$--the contribution of
each data point to the $\chi_{\rm min}^2$.}
\label{tab:dchi2}
\end{table}

Until now we have assumed exact $SU(3)$ flavor symmetry.
Before considering NP as a solution of these $B\to \pi K$
puzzles,
we proceed to improve the SM parametrization by including smaller
amplitudes we have neglected in the above analysis and/or
by taking SU(3) breaking effects into account.
First we include $P'_{uc}$ and $P^{'C}_{\rm EW}$ which
are subdominant according to (\ref{eq:hierarchy}) in the
decay amplitudes. Then we the decay amplitudes of $B \to \pi\pi$
and $B \to \pi K$ are corrected to be
\bea
\sqrt{2} A(B^+ \to \pi^+ \pi^0) &=& -\l(T + C \r)  e^{i\gamma},
\nl
A(B^0 \to \pi^+ \pi^-) &=& -\l(T e^{i\gamma} + P e^{-i\beta} \r), \nl
\sqrt{2} A(B^0 \to \pi^0 \pi^0) &=& -C^{i\gamma} + P e^{-i\beta}, \nl
A(B^+ \to \pi^+ K^0) &=& -P' -{1 \over 3} P^{'C}_{\rm EW}
+ P^{'}_{uc} e^{i\gamma} , \nl
\sqrt{2} A(B^+ \to \pi^0 K^+) &=& P' - P'_{\rm EW}
-{2 \over 3} P^{'C}_{\rm EW} - \l(T' + C' + P^{'}_{uc} \r) e^{i\gamma} , \nl
A(B^0 \to \pi^- K^+) &=& P' -{2 \over 3} P^{'C}_{\rm EW} -\l(T' + P^{'}_{uc}\r) e^{i\gamma}, \nl
\sqrt{2} A(B^0 \to \pi^0 K^0) &=&
-P' - P'_{\rm EW}  -{1 \over 3} P^{'C}_{\rm EW}
-\l(C' - P^{'}_{uc}\r) e^{i\gamma}.
\label{eq:all2}
\eea
We also incorporate the factorizable $SU(3)$ breaking effect
to the tree amplitude~\cite{Rosner:2004} so that
\bea
 {T \over T'} = {f_\pi \over f_K} \; {V_{ud} \over V_{us}},
\eea
where $f_\pi$ $(f_K)$ is the decay constant of $\pi$ $(K)$.
For numerical analysis we used $f_{\pi(K)}= 131 (160)$ (MeV).
For color-suppressed tree and QCD penguin amplitudes we still use the
$SU(3)$ relation (\ref{eq:su3}). We also
use the relation (\ref{eq:GPY}) for electroweak penguins in terms
of trees ({\bf SM fit III}).

As can be seen
in Table~\ref{tab:SM-fit}, the $\chi^2_{\rm min}$ does not improve
at all. In addition $|P'_{uc}|$ which should be much smaller compared
with $|T'|$ does not follow this hierarchy. We also see that
$\Delta \chi^2_{\rm min}$'s for $S_{\rm CP}(B^0 \to \pi^0 K^0)$
and $A_{\rm CP}(B^+ \to \pi^0 K^+)$
in Table~\ref{tab:dchi2}
are still troublesome.
Therefore we conclude that the inclusion of factorizable $SU(3)$ breaking
effect in $T^{(')}$ and smaller amplitudes  $P'_{uc}$ and $P^{'C}_{\rm EW}$ alone
does not help improving the SM fit.

Now we consider the effect of reasonable $SU(3)$ breaking.
To do this we introduce two parameters $b_C$ and $b_P$ to represent
the $SU(3)$ breaking for the color-suppressed tree and QCD penguin so that
\bea
 {C \over C'} = b_C \; {V_{ud} \over V_{us}}, \quad
 {P \over P'} = b_P \; \l|V_{td} \over V_{ts} \r|.
\label{eq:su3_br}
\eea
We added two free parameters $b_C$ and $b_P$ to ``SM fit III'' ({\bf SM fit IV})
and obtained $b_C = 3.5 \pm 6.6$ and $b_P = 1.7 \pm 0.7$ ($\chi^2_{\rm min}/dof=4.7/5$).
Although they have huge errors, the central values require too large
$SU(3)$ breaking effect, considering the fact that it is expected to be at most $20-30\%$.
To make matters worse, not only $|P'_{uc}|$ is too large but
$\gamma=41^\circ \pm 5^\circ$ is
much lower than that obtained in the global CKM fitting~\cite{CKMfitter}.

\section{NP fitting}
\label{sec:NP}

\begin{table}[tb]
\center
\begin{tabular}{cccc}
\hline
\hline
 & NP fit I   & NP fit II & NP fit III\\
\hline
$\chi_{\rm min}^2/dof~(\mbox{quality of fit})$ & $6.28/7~(51\%)$
& $7.92/7~(34\%)$ & $8.3/7~(31\%)$\\
\hline
$\gamma$ & $71.7^\circ \pm 5.7^\circ$ & $71.1^\circ \pm 8.4^\circ$
         & $53.2^\circ \pm 8.7^\circ$ \\
$|T'|$~(eV) & $5.21 \pm 0.27$ & $5.23 \pm 0.28$
         & $5.40 \pm 0.30$ \\
$\delta_{T'}$ & $30.3^\circ \pm 5.5^\circ$ & $32.9^\circ \pm 13.6^\circ$
         & $75.0^\circ \pm 40.9^\circ$ \\
$|C'|$~(eV) & $3.25 \pm 0.54$ & $3.56 \pm 0.54$
         & $4.41 \pm 0.43$ \\
$\delta_{C'}$ & $-14.5^\circ \pm 18.7^\circ$ & $-24.4^\circ \pm 15.4^\circ$
              & $-0.4^\circ \pm 35.8^\circ$ \\
$|P'|$~(eV) & $48.6 \pm 0.7$ & $47.4 \pm 4.3$
            & $25.2 \pm 8.0$\\
%$\delta_{P'}$ & 0 (fixed) & 0 (fixed)\\
$\delta_{\rm NP}$ & $7.6^\circ \pm 4.3^\circ$ & $-5^\circ \pm 2^\circ$ & $-100^\circ \pm 44^\circ$\\
$|P'_{\rm EW,NP}|$~(eV) & $20.1 \pm 4.7$ & - & - \\
$\phi_{\rm EW}$~(eV) & $-87.4^\circ \pm 4.5^\circ$ & - & - \\
$|P^{'C}_{\rm EW,NP}|$~(eV) & - & $33.6 \pm 26.7$ & - \\
$\phi^C_{\rm EW}$~(eV) & -& $-88^\circ \pm 3^\circ$ & - \\
$|P'_{\rm NP}|$~(eV) & - & - & $39.0\pm 19.3$ \\
$\phi_{\rm P}$~(eV) & - & - & $-1.94^\circ \pm 2.12^\circ$ \\
\hline
\hline
\end{tabular}
\caption{Results for ``NP fit I'', ``NP fit II'' and ``NP fit III''.
See the text for details.}
\label{tab:NP-fit}
\end{table}

\begin{table}[t]
\center
\begin{tabular}{cccc}
\hline
\hline
Observable & NP fit I  & NP fit II & NP fit III\\
\hline
$BR(B^+ \to \pi^+ \pi^0)$ & $3.5\times 10^{-4}$  & 0.071 & 0.037\\
$BR(B^0 \to \pi^+ \pi^-)$ & 0.056 & 0.086 & 0.030\\
$BR(B^0 \to \pi^0 \pi^0)$ & $2.4 \times 10^{-7}$ & 0.21 & 0.032\\
$BR(B^+ \to \pi^+ K^0)$ & 1.7 & 0.13 & 0.089\\
$BR(B^+ \to \pi^0 K^+)$ & 0.16 & 0.29 & 0.70\\
$BR(B^0 \to \pi^- K^+)$ & 1.1 &  0.026 & 0.97\\
$BR(B^0 \to \pi^0 K^0)$ & 0.099 & 0.46 & 1.2\\
\hline
$A_{\rm CP}(B^+ \to \pi^+ \pi^0)$ & 0.028 & 0.028 &0.028\\
$A_{\rm CP}(B^0 \to \pi^+ \pi^-)$ & 0.14 0 & 0.31 & 0.023\\
$A_{\rm CP}(B^0 \to \pi^0 \pi^0)$ & 0.021 & 0.037 & 0.48\\
$A_{\rm CP}(B^+ \to \pi^+ K^0)$ & 0.25 & 2.0 & 0.024\\
$A_{\rm CP}(B^+ \to \pi^0 K^+)$ & 0.13 & 0.42 & 0.15\\
$A_{\rm CP}(B^0 \to \pi^- K^+)$ & 0.16 & 0.29 & 0.097\\
$A_{\rm CP}(B^0 \to \pi^0 K^0)$ & 1.8 & 0.0013 & 1.4\\
\hline
$S_{\rm CP}(B^0 \to \pi^+ \pi^-)$ & 0.18 & 0.058 & 0.042\\
$S_{\rm CP}(B^0 \to \pi^0 K^0)$ & 0.49 & 3.5 & 2.9\\
\hline
\hline
\end{tabular}
\caption{$\Delta \chi_{\rm min}^2$--the contribution of
each data point to the $\chi_{\rm min}^2$.}
\label{tab:dchi2_NP}
\end{table}

We have seen in Section~\ref{sec:SM} that the SM parametrization does
not describe the experimental data very well.
Although the discrepancy is about $2-3 ~\sigma$ level
and we cannot rule out the
SM yet, it would be interesting to investigate whether a new parametrization
coming from NP will improve the fitting.

Since the parameterizations in (\ref{eq:all}) and (\ref{eq:all2})
can perfectly fit to the $B\to \pi\pi$ data,
we will assume that NP appears only in the $B \to \pi K$ modes.
A given NP model can generate many new terms in the decay amplitudes
with their own weak phases and strong phases.
To simplify the analysis we adopt a reasonable argument that
the strong phases of NP are negligible~\cite{Baek:piK}.
With this assumption it is enough to introduce just one NP amplitude
with effective weak phase for each topological amplitude.
We assume there is no NP contribution to tree amplitude $T'$
and color-suppressed tree amplitude $C'$.
Then the decay amplitudes can be written as
\bea
\sqrt{2} A(B^+ \to \pi^+ \pi^0) &=& -\l(T + C \r)  e^{i\gamma},
\nl
A(B^0 \to \pi^+ \pi^-) &=& -\l(T e^{i\gamma} + P e^{-i\beta} \r), \nl
\sqrt{2} A(B^0 \to \pi^0 \pi^0) &=& -C^{i\gamma} + P e^{-i\beta}, \nl
A(B^+ \to \pi^+ K^0) &=& -P' + P'_{\rm NP} e^{i \phi_P}
-{1 \over 3} P^{'C}_{\rm EW,NP} e^{i \phi^C_{\rm EW}}, \nl
\sqrt{2} A(B^+ \to \pi^0 K^+) &=& P' - T'e^{i\gamma}  - C'e^{i\gamma}
 - P'_{\rm EW}
 - P'_{\rm NP} e^{i \phi_P}
 - P'_{\rm EW,NP} e^{i \phi_{\rm EW}} \nl
&& - {2 \over 3} P^{'C}_{\rm EW,NP} e^{i \phi^C_{\rm EW}}
, \nl
A(B^0 \to \pi^- K^+) &=& P' - T'e^{i\gamma}  - P'_{\rm NP} e^{i \phi_P}
 - {2 \over 3} P^{'C}_{\rm EW,NP} e^{i \phi^C_{\rm EW}}, \nl
\sqrt{2} A(B^0 \to \pi^0 K^0) &=&
-P' - C' e^{i\gamma} -P'_{\rm EW}
 + P'_{\rm NP} e^{i \phi_P}
 - P'_{\rm EW,NP} e^{i \phi_{\rm EW}} \nl
&& - {1 \over 3} P^{'C}_{\rm EW,NP} e^{i \phi^C_{\rm EW}}
.
\label{eq:NPall}
\eea
Note that we included NP contribution to the color-suppressed electroweak
diagram. This is because this contribution need not be suppressed
compared to the electroweak penguin while it is
actually suppressed in the SM.

This description of NP has 7 additional parameters, overall
strong phase $\delta_{\rm NP}$ relative to that of $P'$ which we
set to be zero,
three real NP amplitudes and three NP weak phases. Using all the
new parameters in fitting makes statistics quite poor.
So, at first, we assume one NP terms dominate and neglect the others.

First, we consider only the effect of $P'_{\rm EW}$ ({\bf NP fit I})
which corresponds
to the solution considered in~\cite{Buras:EW}.
Table~\ref{tab:NP-fit} shows that we obtained an excellent fit
for this scenario.
In addition, we can see in Table~\ref{tab:dchi2_NP} that all the puzzling
behaviors
of the SM have disappeared. The largest deviation from the best fit parameters
is at most $1.3 \sigma$.

Now we consider a scenario where $P^{'C}_{\rm EW, NP}$ dominates
({\bf NP fit II}).
We can see from Table~\ref{tab:NP-fit} that this fit is also acceptable.
However, the data for $S_{\rm CP}(B^0 \to \pi^0 K^0)$ is a little bit away
from the best fitted values.

Similarly we can consider the case where only $P'_{\rm NP}$ exists
({\bf NP fit III}). In this case,
although $\chi^2_{\rm min}/dof$ is acceptable,
$\Delta \chi^2_{\rm min}$ of $S_{\rm CP}(B^0 \to \pi^0 K^0)$ is not so satisfactory.

%\begin{table}[tb]
\begin{table}
\center
\begin{tabular}{cccc}
\hline
\hline
 & NP fit IV   & NP fit V & NP fit VI\\
\hline
$\chi_{\rm min}^2/dof~(\mbox{quality of fit})$ & $2.45/5~(78.0\%)$  & $4.55/5~(47\%)$ & $0.51/5~(99\%)$\\
                                               & $2.72/5~(74.4\%)$ &  & $0.97/5~(97\%)$\\
\hline
$\gamma$ & $69.7^\circ \pm 5.6^\circ$ & $62.4^\circ \pm 8.4^\circ$ & $60.7^\circ \pm 8.8^\circ$ \\
         & $75.5^\circ \pm 5.6^\circ$ &                            & $55.0^\circ \pm 7.1^\circ$ \\
$|T'|$~(eV) & $5.26 \pm 0.24$ & $5.25 \pm 0.24$ & $5.27 \pm 0.24$ \\
            & $5.29 \pm 0.28$ &                 & $5.35 \pm 0.27$ \\
$\delta_{T'}$ & $31.9^\circ \pm 9.0^\circ$ & $43.1^\circ \pm 15.1^\circ$ & $50.4^\circ \pm 23.3^\circ$ \\
              & $26.5^\circ \pm 7.3^\circ$ &                             & $70.5^\circ \pm 27.2^\circ$ \\
$|C'|$~(eV) & $3.71 \pm 0.50$ & $4.03 \pm 0.58$  & $4.15 \pm 0.51$ \\
            & $3.02 \pm 0.54$ &                  & $4.35 \pm 0.44$ \\
$\delta_{C'}$ & $-27.7^\circ \pm 11.1^\circ$ & $-21.4^\circ \pm 16.5^\circ$ & $-17.1^\circ \pm 18.9^\circ$ \\
              & $-13.8^\circ \pm 20.3^\circ$ &                              & $-2.36^\circ \pm 22.7^\circ$ \\
$|P'|$~(eV) & $40.0 \pm 2.6$ & $32.6 \pm 11.9$  & $30.6 \pm 11.0$\\
            & $49.6 \pm 0.9$ &                  & $26.2 \pm  8.3$\\
%$\delta_{P'}$ & 0 (fixed) & 0 (fixed)\\
$\delta_{\rm NP}$ & $177^\circ \pm 1^\circ$ & $9.04^\circ \pm 6.95^\circ$ & $-1.59^\circ \pm 4.60^\circ$\\
                  & $7.9^\circ \pm 4.8^\circ$ &                           & $-179^\circ \pm 5^\circ$\\
$P'_{\rm EW,NP}$~(eV) & $62.2 \pm 6.9$ & $21.1 \pm 4.7$ & - \\
                        & $19.4 \pm 6.0$ &   & - \\
$\phi'_{\rm EW}$ & $-75.9^\circ \pm 3.8^\circ$ & $-87.2^\circ \pm 6.0^\circ$ & - \\
                     & $-90.9^\circ \pm 5.0^\circ$ &  & - \\
$P^{'C}_{\rm EW,NP}$~(eV) & $65.4 \pm 7.4$ & - & $86.4 \pm 19.5$ \\
                            & $3.96 \pm 4.34$ &  & $-61.7 \pm 21.1$ \\
$\phi^{'C}_{\rm EW}$ & $106^\circ \pm 3^\circ$ & - & $146^\circ \pm 23^\circ$ \\
                       & $31.6^\circ \pm 78.3^\circ$ &  & $27.7^\circ \pm 15.4^\circ$ \\
$P'_{\rm NP}$~(eV) & - & $16.9\pm 12.2$ & $56.7\pm 13.3$ \\
                     & - &                & $25.3\pm 5.0$ \\
$\phi'_{\rm P}$ & - & $-167^\circ \pm 16^\circ$ & $6.91^\circ \pm 6.93^\circ$ \\
                     & - &                           & $-81.5^\circ \pm 23.6^\circ$ \\
\hline
\hline
\end{tabular}
\caption{The fits for NP (IV, V, VI). See the text for details.}
\label{tab:NP-fit2}
\end{table}

\begin{table}
\center
\begin{tabular}{ccccccc}
\hline
\hline
\multicolumn{7}{c}{NP fit VII}\\
\hline
\hline
$\chi_{\rm min}^2/dof$ & $\gamma$ & $|T'|$ & $\delta_{T'}$ & $|C'|$ & $\delta_{C'}$ & $|P'|$  \\
\hline
0.21(98\%)      & $63 \pm 10$ & $5.3 \pm 0.2$ & $44 \pm 23$ & $4.1 \pm 0.6$ & $-23 \pm 19$ & $33 \pm 12$\\
0.43(93\%)      & $73 \pm 8$  & $5.2 \pm 0.3$ & $73 \pm 8$  & $3.3 \pm 0.8$ & $-15 \pm 21$ & $46 \pm 12$\\
2.6 (45\%)      & $69 \pm 31$ & $5.2 \pm 0.3$ & $33 \pm 41$ & $3.6 \pm 2.5$ & $-20 \pm 25$ & $41 \pm 41$\\
\hline
\hline
$\delta_{\rm NP}$ & $P'_{\rm EW,NP}$ & $\phi'_{\rm EW}$ & $P^{'C}_{\rm EW,NP}$
  & $\phi^{'C}_{\rm EW}$ & $P'_{\rm NP}$ & $\phi'_{\rm P}$ \\
\hline
$179 \pm 2$ & $2.1 \pm 3.8$ & $9.8 \pm 52$ & $76 \pm 21$ & $-93 \pm 22$ &  $15 \pm 8$ & $104 \pm 37$ \\
$0.6 \pm 1.3$ & $65 \pm 45$ & $33 \pm 29$ & $65 \pm 44$ & $-140 \pm 30$ & $48 \pm 24$ & $48 \pm 38$ \\
$-172 \pm 8$ & $19 \pm 9$   & $89 \pm 7$  & $2.7 \pm 3.9$ & $-180 \pm 340$ & $8.6 \pm 42$ & $5.7 \pm 38$ \\
\hline
\hline
\end{tabular}
\caption{The fit for ``NP fit VII''. }
\label{tab:all_NP}
\end{table}

We can analyze more general cases of having two-types of NP
simultaneously. We have 11 parameters to fit in these cases.
For each case we obtained several acceptable
solutions, which is due to the low statistics.
With only $P'_{\rm EW, NP}$ and $P^{'C}_{\rm EW, NP}$ ({\bf NP fit IV})
we get two distinctive
solutions in Table~\ref{tab:NP-fit2}.
It is interesting to note that the best solution of ``NP fit IV''
favors the enhancement of both $P'_{\rm EW, NP}$ and
 $P^{'C}_{\rm EW, NP}$ contrary to Ref.~\cite{Buras:EW}.
As mentioned earlier,  $P^{'C}_{\rm EW, NP}$ is not necessarily
color suppressed and can be as large as  $P'_{\rm EW, NP}$.
The second solution corresponds to that found in Ref.~\cite{Buras:EW},
{\it i.e.} NP in the electroweak penguin sector.

The scenario of having non-vanishing $P'_{\rm NP}$ and
$P'_{\rm EW, NP}$ ({\bf NP fit V})  shows that
$P'_{\rm NP}$ need not be suppressed and can be almost as large
as $P'_{\rm EW, NP}$.
Even when $P'_{\rm EW, NP}$ vanishes ({\bf NP fit VI}), the
large contribution of $P'_{\rm NP}$ and $P^{'C}_{\rm EW, NP}$
can solve the $B \to \pi K$ puzzle.

Now we consider
the simultaneous contribution of all the possible NP
contributions, {\it i.e.} $P'_{\rm NP}$,$P'_{\rm EW,NP}$ and
$P^{'C}_{\rm EW,NP}$ for completeness ({\bf NP fit VII}),
although we have poor statistics for definite prediction.
As expected, we obtained many
physically acceptable local minima. We list just three of them in Table~\ref{tab:all_NP}.
Since the $\chi^2_{\rm min}$'s of ``NP fit VI'' are quite low,
the solutions  with low-lying $\chi^2_{\rm min}$
values look similar to those of  ``NP fit VI'' (See the 1st solution).
The 2nd solution shows that all the 3 types of NP can be sizable.
The 3rd solution corresponds to the ``NP fit I'' which is the NP in the electroweak penguin
sector obtained in~\cite{Buras:EW}.

\section{Conclusions}
\label{sec:con}
We performed $\chi^2$ fitting to check if the conventional parametrization in the SM
describes well the current experimental data of $B \to \pi \pi$ and $B \to \pi K$
decays.
Contrary to the data used in Ref.~\cite{Rosner:2004}, the current data disfavors this
parametrization given in (\ref{eq:all}) and (\ref{eq:all2}) at 2--3 $\sigma$ level.

We interpreted this difficulty in the SM as a manifestation of NP and investigated various NP solutions.
When a single NP amplitude dominates, NP in the electroweak penguin sector is the most
favorable solution in accord with~\cite{Buras:EW}.
When two or more NP amplitudes exist simultaneously, solutions other than
in the electroweak penguin sector can also explain the
deviation very well.

\vskip1.3cm
\noindent {\bf Acknowledgment}\\
The author thanks  C.~S.~Kim for useful comments
and P. Ko for warm hospitality during his visit to KIAS where part of this work
was done.

\end{document}